\documentclass[pre,showpacs,twocolumn,floatfix,amsmath,amsfonts]{revtex4}

\newcommand{\cX}{\mathcal{X}}
\newcommand{\g}{\gamma}
\newcommand{\cY}{\mathcal{Y}}
\newcommand{\dl}{\delta_k}
\newcommand{\be}{\beta}
\newcommand{\E}{\mathrm{e}}
\newcommand{\iu}{\mathrm{i}}
\newcommand{\cZ}{\mathcal{Z}}

\DeclareMathOperator*{\res}{res} \DeclareMathOperator{\IM}{Im}

\usepackage{epsfig}
\usepackage{graphicx}
\usepackage{amsmath}
\usepackage{amssymb}
\usepackage{dcolumn}

\usepackage{dsfont, srcltx}

\begin{document}

\title{Kink scattering from a parity-time-symmetric defect in the $\phi^4$ model}

\author{Danial~Saadatmand$^{1}$}
\email{saadatmand.d@gmail.com}

\author{Sergey~V.~Dmitriev$^{2,3}$}
\email{dmitriev.sergey.v@gmail.com}

\author{Denis~I.~Borisov$^{4,5}$}
\email{borisovdi@yandex.ru}

\author{Panayotis~G.~Kevrekidis$^{6}$}
\email{kevrekid@math.umass.edu}

\author{Minnekhan~A.~Fatykhov$^{5}$}
\email{fatykhovma@mail.ru}

\author{Kurosh~Javidan $^{1}$}
\email{javidan@um.ac.ir}

\affiliation{ $^1$Department of Physics, Ferdowsi University of
Mashhad, 91775-1436 Mashhad, Iran
\\
$^2$Institute for Metals Superplasticity Problems RAS, Khalturin
39, 450001 Ufa, Russia
\\
$^3$National Research Tomsk State University, Lenin Prospekt 36,
634050 Tomsk, Russia
\\
$^4$Institute of Mathematics CC USC RAS, Chernyshevsky 112, 450008
Ufa, Russia
\\
$^5$Bashkir State Pedagogical University, October Revolution St.
3a, 450000 Ufa, Russia
\\
$^6$Department of Mathematics and Statistics, University of
Massachusetts, Amherst, MA 01003 USA }

\begin{abstract}
In this paper, we study the $\phi^4$ kink scattering from a
spatially localized $\mathcal{PT}$-symmetric defect and the effect
of the kink's internal mode (IM) is discussed. It is demonstrated
that if a kink hits the defect from the gain side, a noticeable IM
is excited, while for the kink coming from the opposite direction
the mode excitation is much weaker. This asymmetry is a principal
finding of the present work. Similar to the case of the
sine-Gordon kink studied earlier, it is found that the $\phi^4$
kink approaching the defect from the gain side always passes
through the defect, while in the opposite case it must have
sufficiently large initial velocity, otherwise it is trapped by
the loss region. It is found that for the kink with IM the
critical velocity is smaller, meaning that the kink bearing IM can
pass {\it more easily} through the loss region. This feature,
namely the ``increased transparency'' of the defect as regards the
motion of the kink in the presence of IM is the second key finding
of the present work. A two degree of freedom collective variable
model offered recently by one of the co-authors is shown to be
capable of reproducing both principal findings of the present
work. A simpler, analytically tractable single degree of freedom collective variable method is
used to calculate analytically the kink phase shift and the kink
critical velocity sufficient to pass through the defect.
Comparison with the numerical results suggests that the collective
variable method is able to predict these parameters with a high
accuracy.
\end{abstract}
\pacs {05.45.Yv, 45.50.Tn} \maketitle

\section {Introduction}

About 15 yeas ago the standard notion of Hermiticity in
quantum mechanics was challenged by the
studies of Bender and co-authors, who demonstrated that a class
of non-Hermitian Hamiltonians possess real spectra under the
parity-time ($\mathcal{PT}$) symmetry condition, where parity-time
means spatial reflection and time reversal, $x\rightarrow-x$ and
$t\rightarrow-t$ \cite{Bender1,Bender2}. This mathematical
discovery has generated an intense interest
in the consideration of open physical systems
with balanced gain and loss and such systems have been realized
experimentally in optics
\cite{Ruter,Guo,Regensburger,Regensburger1,Peng2014,peng2014b}, electronic
circuits \cite{Schindler1,Schindler2,Factor}, and mechanical
systems \cite{Bender3}.

$\mathcal{PT}$-symmetric systems have attracted a great deal of
attention from researchers in different branches within physics
because they can demonstrate unusual and counterintuitive
properties such as unconventional beam refraction
\cite{Zheng010103}, Bragg scattering \cite{Longhi022102},
symmetry-breaking transitions \cite{Ruter} and associated ghost
states~\cite{wunner,graefe,usbender,vassos}, a loss-induced
optical transparency \cite{Guo}, conical diffraction
\cite{Ramezani013818}, a new type of Fano resonance
\cite{Miroshnichenko012123}, chaos \cite{West054102}, nonlocal
boundary effects \cite{Sukhorukov2148}, optical switches
\cite{Nazari} and diodes~\cite{jennie1,jennie2}, phase sensitivity
of light dynamics \cite{Barashenkov,Suchkov,Rysaeva}, and the
possibility of linear and nonlinear wave amplification and
filtering~\cite{Dmitriev013833,Suchkov033825,SuchkovEPL}. 1D and
2D $\mathcal{PT}$-symmetric discrete models can demonstrate
unexpected instabilities
\cite{kaili,barflach2,pickton,dep1,dep2,kondep,uwe}. Extensions of
$\mathcal{PT}$-symmetric considerations in the setting of active
media with not perfectly balanced gain and loss have also recently
been proposed~\cite{DSK2976,barflach,haitao}. The possibility of a
Hamiltonian reformulation of the standard $\mathcal{PT}$-symmetric
dimer has been recently demonstrated \cite{Barashenkov1} and calls
for the broader question of the potential Hamiltonian/Hermitian
nature of such systems in suitable modified variables.

Motivated by the linear oscillator problems associated with the
$\mathcal{PT}$-symmetric electrical~\cite{Schindler1,Schindler2}
and mechanical~\cite{Bender3} experiments, one of the authors has
recently generalized the Klein-Gordon field theory by introducing
a $\mathcal{PT}$-symmetric terms within it~\cite{KevrekidisRevA},
including the case example of a localized $\mathcal{PT}$-symmetric
defect. A collective coordinate method was developed in this work
to describe the kink interaction with the defect, see
also~\cite{Demirkaya2,Demirkaya,PanosNew}. It was shown that
standing kinks in such models are stable if they are centered at
the loss side of the defect~\cite{Demirkaya} (and unstable on the
gain side), while standing breathers may exist only if centered
exactly at the interface between gain and loss regions
\cite{PanosNew}.

It should be pointed out that the interaction of solitary waves
with each other \cite{S1,S2,S3,S4} and with local inhomogeneities
of media has been analyzed in various settings. E.g., early works
considered the interaction of a fluxon with a non-conservative
localized inhomogeneity in a long Josephson junction in
\cite{KivsharM,KivsharMN}. The excitation of both the kink's
internal mode (IM) and the defect mode, due to the collision of
the kink with a local defect, has been described \cite{Boris}. The
reflection windows were observed in the kink-impurity interactions
in the sine-Gordon (SG) \cite{Fei1} and $\phi^4$ \cite{Fei2}
models. Scattering of the SG breather by localized defects has
been investigated \cite{Piette}. Scattering of linear and
nonlinear waves (solitons) on defects in $\mathcal{PT}$-symmetric
optical waveguide arrays was analyzed
\cite{Dmitriev013833,Suchkov033825,SuchkovEPL,jennie1,jennie2}. It
was shown that the incident high-amplitude solitons (or even
linear wavepackets~\cite{jennie1,jennie2}) can excite a mode
localized on the $\mathcal{PT}$-symmetric defect. Scattering of
wavepackets in such systems was shown to depend on the direction
of incidence.

Recently, the interaction of the moving kinks and breathers with
the spatially localized $\mathcal{PT}$-symmetric perturbation was
investigated in the realm of the SG field \cite{Danial}. Several
new soliton-defect interaction scenarios were observed such as the
kink passing/trapping depending on whether the kink comes from the
gain or loss side of the impurity, merger of the kink-antikink
pair into breather, and splitting of the breather into
kink-antikink pair. The collective variable method
\cite{KevrekidisRevA} was successfully applied to calculate the
kink phase shift as a result of interaction with the impurity and
the threshold kink velocity to pass through the lossy side of the
 defect.

The SG kink does not support a vibrational IM, while the kinks in
the non-integrable $\phi^4$ model do support such a
mode~\cite{KivsharIM,Quintero}. It is for that reason that the
kink-antikink interactions are far richer in the case of the
$\phi^4$ model~\cite{Campbell2,Belova}; see also for a recent
discussion~\cite{haberman}. When a kink hits an impurity in a
conservative model, a part of its energy is trapped towards the
excitation of the impurity mode \cite{Fei1,Kivshar1177} and
another fraction leads to the emission of radiation bursts
\cite{Malomed385}. It is of particular interest to investigate the
role of the kink's IM in the case when the kink interacts with the
$\mathcal{PT}$-symmetric impurity. This problem is addressed here
for the $\phi^4$ kinks.

The structure of the paper is as follows. In Sec.~\ref{Sec:II},
following the work \cite{KevrekidisRevA}, we introduce the
spatially localized $\mathcal{PT}$-symmetric inhomogeneity into
the $\phi^4$ field and present the well-known $\phi^4$ kink
solution and the kink's IM profile. In Sec.~\ref{Sec:CollectVar},
a collective variable method is applied and analytically solved to
reveal some features of the kink dynamics in the considered
system. We report on the numerical results for scattering of kinks
on the $\mathcal{PT}$-symmetric defect in Sec.~\ref{Sec:IIA}. Our
conclusions and some future directions are presented in
Sec.~\ref{Sec:V}. A number of technical details on the analytical
calculations are presented in the Appendix.

\section {The model} \label{Sec:II}
We consider the modified $\phi^4$ equation of the form
\cite{KevrekidisRevA}
\begin{equation}\label{phi4a}
a\phi_{tt} - b\phi_{xx} -c\phi(1-\phi^2) = d\gamma(x)\phi_{t},
\end{equation}
where $\phi(x,t)$ is the unknown scalar field, lower indices
denote partial differentiation and $a$, $b$, $c$, and $d$ are the
coefficients. In terms of new variables $t\rightarrow
t\sqrt{2a/c}$ and $x\rightarrow x\sqrt{2b/c}$, Eq.~(\ref{phi4a})
assumes the dimensionless form
\begin{equation}\label{phi4}
\phi_{tt} - \phi_{xx} -2\phi(1-\phi^2) =
\epsilon\gamma(x)\phi_{t},
\end{equation}
in which parameter $\epsilon=\sqrt{2d^2/ac}$ controls the defect
amplitude. In order to study the effects of a spatially localized
$\mathcal{PT}$-symmetric defect on traveling kinks, for the
function $\gamma(x)$ we take
\begin{equation}\label{Phi4_Perturbation}
\gamma(x)=\tanh(\beta x){\rm sech}(\beta x),
\end{equation}
which has the symmetry $\gamma(-x)=-\gamma(x)$. Physically this
implies that while Eq.~(\ref{phi4}) describes an open system with
gain and loss, the gain balances the loss. Parameter $\beta$
characterizes the defect inverse width.

For $\gamma(x)\equiv 0$, we have the non-integrable $\phi^4$
equation with the following moving kink solution
\begin{equation}\label{Kink}
   \phi_K(x,t)=\pm\tanh\{\delta_k(x-x_0-V_k t)\},
\end{equation}
where $V_k$ is the kink velocity, $x_0$ is the kink initial
position and $\delta_{k} =1/\sqrt{1 - V_{k}^2}$. The energy of the
kink is $E_k=4\delta_{k}/3$.

An approximate solution (i.e., a solution to leading order in the
linearization amplitude $A$) for the kink bearing IM can be
presented in the form \cite{Belova}
\begin{equation}\label{KinkModeSolution}
  \Phi_K(x,t)=\phi_K(x,t)+A\xi(x,t),
\end{equation}
with the kink's IM mode profile
\begin{equation}\label{KinkMode}
   \xi(x,t)=\sqrt{\frac{3}{2}}\tanh\{\delta_k(x-x_0-V_k t)\}{\rm sech}\{\delta_k(x-x_0-V_k
   t)\}.
\end{equation}
The above Ansatz leads to a kink with the IM having amplitude $A$
and frequency $\omega=\sqrt{3}$. This mode has been discussed by
many authors due to its critical role in the collision
phenomenology of the $\phi^4$ model
\cite{Belova,Campbell2,haberman}. It should be noted, however,
that the above waveform of the so-called ``wobbling kink'' does
not survive indefinitely, but rather decays over time according to
a $t^{-1/2}$ law, as discussed, e.g., in~\cite{bara}; see also
references therein.

As it will be shown, the kink's IM noticeably affects the kink
dynamics during the interaction with the $\mathcal{PT}$-symmetric
defect and it is important from which side the kink hits the
defect.

\section {Collective variable method} \label{Sec:CollectVar}

A two-degree of freedom collective variable model, which takes
into account not only the kink's translational mode but also the
kink's IM has been offered in \cite{KevrekidisRevA}. The $\phi^4$
kink is effectively described by the two degree of freedom
particle of mass $M=4/3$, which is the mass of standing kink. The
kink coordinate $\cX(t)$ (which in the unperturbed case is given
by $x_0+V_k t$ as a function of time $t$) and the kink's internal
mode amplitude $A(t)$ can be found from the following equations
\begin{eqnarray}\label{Coll_var1a}
    M\ddot{\cX}=\epsilon\int_{-\infty}^{\infty} (\phi^{\prime}_K
    +A\xi^{\prime})[(\phi^{\prime}_K
    +A\xi^{\prime})\dot{\cX}-\dot{A}\xi]\gamma(x)dx,
\end{eqnarray}
\begin{eqnarray}\label{Coll_var1aa}
    \ddot{A}=-\omega^2A+\epsilon\int_{-\infty}^{\infty}
    \xi[-(\phi^{\prime}_K
    +A\xi^{\prime})\dot{\cX}+\dot{A}\xi]\gamma(x)dx.
\end{eqnarray}
These equations yield the general form of the nonconservative
forcing including the coupling between the modes. For the case of
weak coupling the above equations simplify as
\begin{eqnarray}\label{Coll_var1}
    M\ddot{\cX}=\epsilon\dot{\cX}\int_{-\infty}^{\infty} [\phi^{\prime}_K
    (x-\cX)]^2\gamma(x)dx,
\end{eqnarray}
\begin{eqnarray}\label{Coll_var11}
    \ddot{A}=-\omega^2A+\epsilon\int_{-\infty}^{\infty}
    [\xi(x-\cX)]^2\gamma(x)dx.
\end{eqnarray}

Below in Sec.~\ref{Sec:CollectVarKV} and
Sec.~\ref{Sec:CollectVarPS} some analytical results for the
simpler single degree of freedom model Eq.~(\ref{Coll_var1}) are
given, while in section \ref{Sec:IIA}, we will present the results
of numerical solution for the two degree of freedom model of
Eqs.~(\ref{Coll_var1a})-(\ref{Coll_var1aa}).

\subsection {Critical kink velocity} \label{Sec:CollectVarKV}

If the kink approaches the defect from the loss side, it must have
sufficient momentum in order to avoid trapping. The critical kink initial
velocity $V_c$ can be found with the help of the collective
variable method. One can present Eq.~(\ref{Coll_var1}) for
$\dot{\cX}$ in the form
\begin{eqnarray}\label{Coll_var4}
    M(\dot{\cX}-\dot{\cX_0})= \epsilon\delta^2_k\int_{-\infty}^{\infty}\int_{\cX_0}^{\cX} \frac{\gamma(x)dxd\cX} {{
    \cosh}^4[\delta_k(x-\cX)]}.
\end{eqnarray}
A kink having critical velocity must have $\dot{\cX}=0$ at
$\cX=0$, i.e., the kink stops when it reaches the center of the
defect. Setting in Eq.~(\ref{Coll_var4}) $V_c=\dot{\cX_0}$ and
$\dot{\cX}=0$ after integrating over the collective variable $\cX$
and $x$ for the initial condition $\cX_0$ and recalling that the
final stopping point is ${\cX}=0$ the critical velocity is given
by the formula
\begin{equation}\label{Coll_var5}
\begin{aligned}
V_c=&\frac{32\epsilon\dl^2}{M}\left( G_1(1)-\E^{4\dl
\cX_0}G_1(\E^{2\dl \cX_0})\right).
\end{aligned}
\end{equation}
The values of $G_1(1)$ and $G_1(\cY)$ can be calculated for the
particular cases $\be=\dl$ and $\be=\dl/2$, as it is shown in
Appendix \ref{AppendixA}. In the former (latter) case the result
is given by Eq.~(\ref{eq14a}) (Eq.~(\ref{eq14aa})). For the kink
with the initial position $\cX_0\rightarrow-\infty$ (far from the
defect), one finds $V_c=32\epsilon\delta^2_kG_1(1)/M$. For small
kink velocity one has $\delta_k \approx 1$ and $M=4/3$ so that
\begin{eqnarray}\label{Vc}
    V_c=0.883\epsilon.
\end{eqnarray}
The above equation indicates that when a slow kink approaches the
loss side of the defect from infinity, its critical velocity $V_c$
is a linear function of the defect amplitude $\epsilon$. Note that
this equation was obtained with the help of the collective
coordinate method Eq.~(\ref{Coll_var1}), accounting solely for the
center of mass variable without taking into account the kink's IM.

\subsection {Kink's phase shift due to interaction with the defect} \label{Sec:CollectVarPS}

The kink approaching the defect from the gain (loss) side is first
accelerated (decelerated) and then decelerated (accelerated) when
it enters the lossy (gain) side. As a result, the kink experiences
a phase shift. The phase shift can be calculated as follows
\cite{Danial}

\begin{equation}\label{Denis19}
\Delta x=\pm\int_{-\infty}^{+\infty}
\frac{\epsilon(F(z)-B)}{\epsilon(F(z)-B)+MV_k}dz,
\end{equation}
where the plus (minus) shows the case when the kink comes from the
loss (gain) side, $V_k>0$ ($V_k<0$).

For the kink solution (\ref{Kink}) the function $F(z)-B$ can be
cast into the particular form
\begin{equation}\label{Denis20}
F(z)-B=-\delta^{-1}_k\int_z^{+\infty}
ds\int\limits_{-\infty}^{+\infty}
\frac{\gamma(x)dx}{\cosh^4[\delta_k(x-s)]}.
\end{equation}
After changing the order of integration and integrating over $s$
one obtains
\begin{align}\label{Denis23}
F(z)-B=-\frac{2}{3\delta_{k} }\int\limits_{-\infty}^{+\infty}&
\gamma(x)\left\{ 1-\tanh[\delta_k(z-x)]\right.\nonumber
\\
&\left.-\frac{\tanh[\delta_k(z-x)]}{2\cosh^2[\delta_k(z-x)]}\right\}dx.
\end{align}
The above integral can be evaluated for the case $\beta=\delta_k$
which is considered below. By evaluating the integral in
Eq.~(\ref{Denis23}) and substituting the result into
Eq.~(\ref{Denis19}) the kink's phase shift can be found as
\begin{align}
&\Delta x=\pm\frac{1}{\dl} G_3\left(-\frac{M V_k \dl^4}{\pi
\epsilon}\right),\label{Denis24a}
\\
&G_3(\cZ):=\int\limits_{0}^{+\infty}
\frac{(y^2+4y+1)dy}{y(y^2+4y+1)-\cZ(y+1)^4},\label{Denis24b}
\end{align}
(see Eq.~(\ref{eq15}) and Eq.~(\ref{eq16}) in the Appendix
\ref{AppendixB}). The phase shift can be calculated only for
$\cZ\in(-\infty,0)\cup(3/8,+\infty)$. For $\cZ\in(0,3/8)$ the kink
hits the defect from the loss side with the velocity smaller than
$V_c$ and is trapped, so that the phase shift diverges. In the
Appendix \ref{AppendixB} the phase shift is calculated explicitly
for the three cases $-\infty<\cZ<-1/8$, $-1/8<\cZ<0$, and
$3/8<\cZ<\infty$.

\section {Numerical Results} \label{Sec:IIA}

To study numerically the effect of the $\mathcal{PT}$-symmetric
defect on the dynamics of the $\phi^4$ kink, we introduce the mesh
$x=nh$, where $h$ is the lattice spacing, $n=0,\pm 1,\pm 2...$ and
propose the following discrete version of Eq.~(\ref{phi4})
\begin{eqnarray}\label{FK_Collisions}
  && \frac{d^2\phi_n}{dt^2} - \frac{1}{h^2}
   \left(\phi_{n-1} - 2\phi_n + \phi_{n+1}\right)+ \nonumber\\
   &&\frac{1}{12h^2} \left(\phi_{n-2} -4\phi_{n-1}+6\phi_n -4\phi_{n+1}+ \phi_{n+2}\right) \nonumber\\
   &&-2\phi_n\left(1-\phi_n^2\right) -\epsilon\gamma_n\frac{d\phi_n}{dt}= 0,
\end{eqnarray}
in which $\phi_n(t)=\phi(nh,t)$ and $\gamma_n=\gamma(nh)$. It can
be seen that the term $\phi_{xx}$ in Eq.~(\ref{phi4}) is
discretized with the accuracy $O(h^4)$ which has already been used
by other authors \cite{BraunKivshar,KivsharMalomed}. This is done
to minimize the effect of discreteness introduced by the mesh.

In order to test the result of the presented analytical method
by means of direct numerical simulations we integrated equations of motion
(\ref{FK_Collisions}) with the help of the St\"{o}rmer method with
respect to the temporal variable using an explicit scheme with the
accuracy of $O(\tau^4)$ and the time step $\tau$. The simulations
were managed for $h=0.1$ and $\tau=0.005$.

To solve numerically the collective variable equations of motion
Eqs.~(\ref{Coll_var1a},\ref{Coll_var1aa}), the temporal variable
is discretized, $t=j\tau$, where $\tau$ is the time step and
$j=0,1,2,...$. The second-order central differences are used to
replace $\ddot{\cX } \sim (\cX _{j-1}-2\cX _j+\cX _{j+1})/\tau^2$,
$\dot{\cX } \sim (\cX _{j+1}-\cX _{j-1})/2\tau$, and similarly for
$\ddot{A}$ and $\dot{A}$. Then
Eqs.~(\ref{Coll_var1a},\ref{Coll_var1aa}) are presented in the
form
\begin{eqnarray}\label{Num1}
  a_{11}\cX _{j+1}+a_{12}A_{j+1}=b_1, \nonumber\\
  a_{21}\cX _{j+1}+a_{22}A_{j+1}=b_2,
\end{eqnarray}
where
\begin{eqnarray}\label{Num2}
  a_{11}&=&\frac{2M}{\tau}-I_1, \quad
  a_{12}=a_{21}=I_2, \quad a_{22}=\frac{2}{\tau}-I_3, \nonumber\\
  b_1&=&\frac{2M}{\tau}(2\cX _j-\cX _{j-1}) - I_1\cX _{j-1} + I_2A_{j-1}, \nonumber\\
  b_2&=&\frac{2}{\tau}(2A_j-A_{j-1}) -2\tau\omega^2A_j - I_3A_{j-1} + I_2\cX _{j-1}, \nonumber\\
  I_1&=&\int_{-\infty}^{\infty} (\phi^{\prime}_K+A_j\xi ^{\prime})^2\gamma(x)dx, \nonumber\\
  I_2&=&\int_{-\infty}^{\infty} (\phi^{\prime}_K+A_j\xi ^{\prime})\xi \gamma(x)dx, \nonumber\\
  I_3&=&\int_{-\infty}^{\infty} \xi ^2\gamma(x)dx.
\end{eqnarray}
To simulate the kink with initial velocity $V_k$ and initial IM
amplitude $A$, for the initial conditions we set $\cX(j=0)=\cX_0$,
$\cX(j=1)=\cX_0+V_k\tau$, $A(j=0)=A(j=1)=A$.

In the present study the simulations are carried out for different
values of the perturbation amplitude $\epsilon$ and fixed
$\beta=1$ (the impurity width is approximately equal to the kink
width). Since we simulate relatively slow kinks with $\delta_k
\approx 1$, the collective variable analytical results,
Eqs.~(\ref{Coll_var5}) and (\ref{Denis24a},\ref{Denis24b}),
calculated for $\beta=\delta_k$ are expected to have a high
accuracy.

\subsection{Kinks bearing no initial IM}

\begin{figure}
\includegraphics[width=8.5cm,height=6cm]{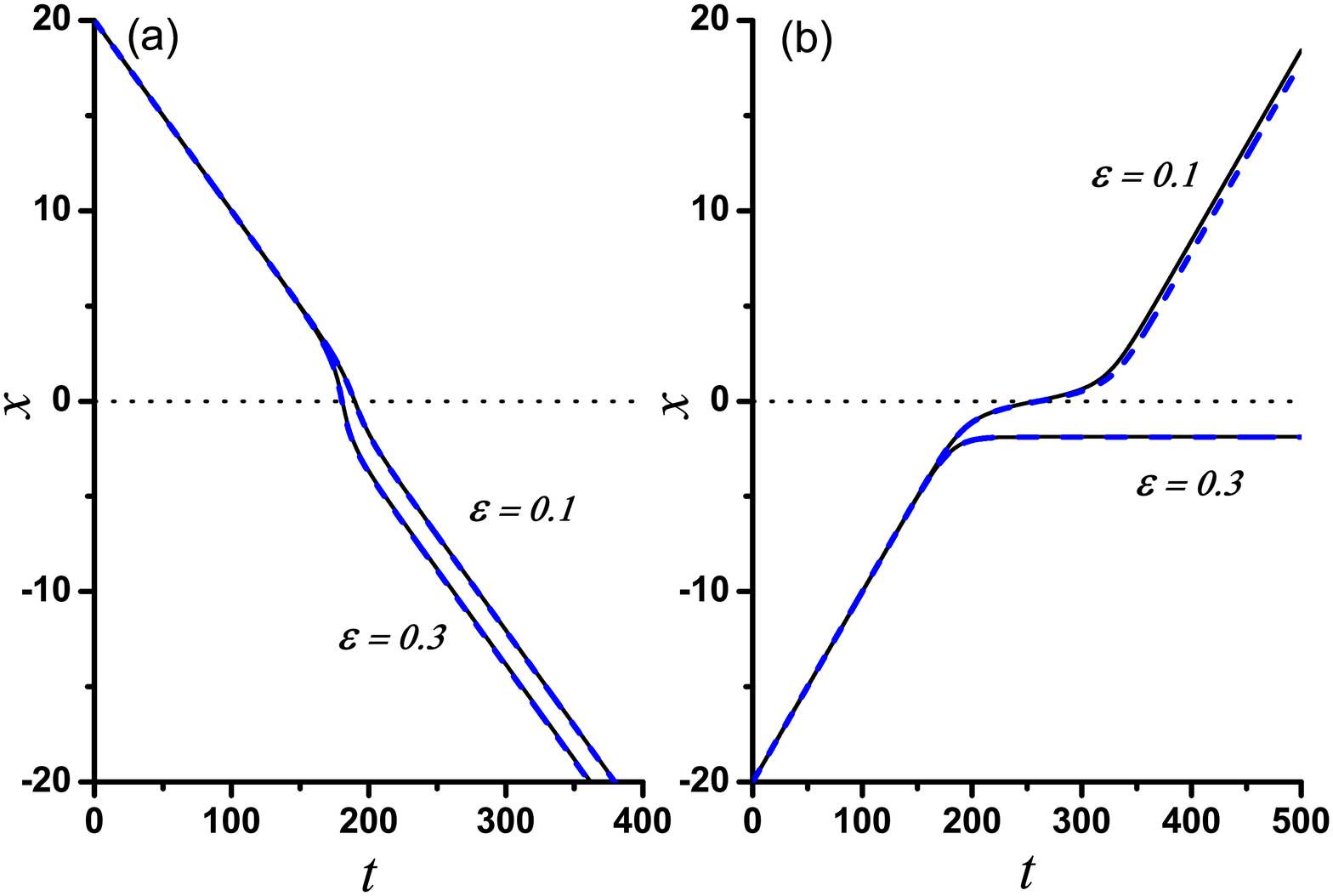}
\includegraphics[width=8.5cm,height=6cm]{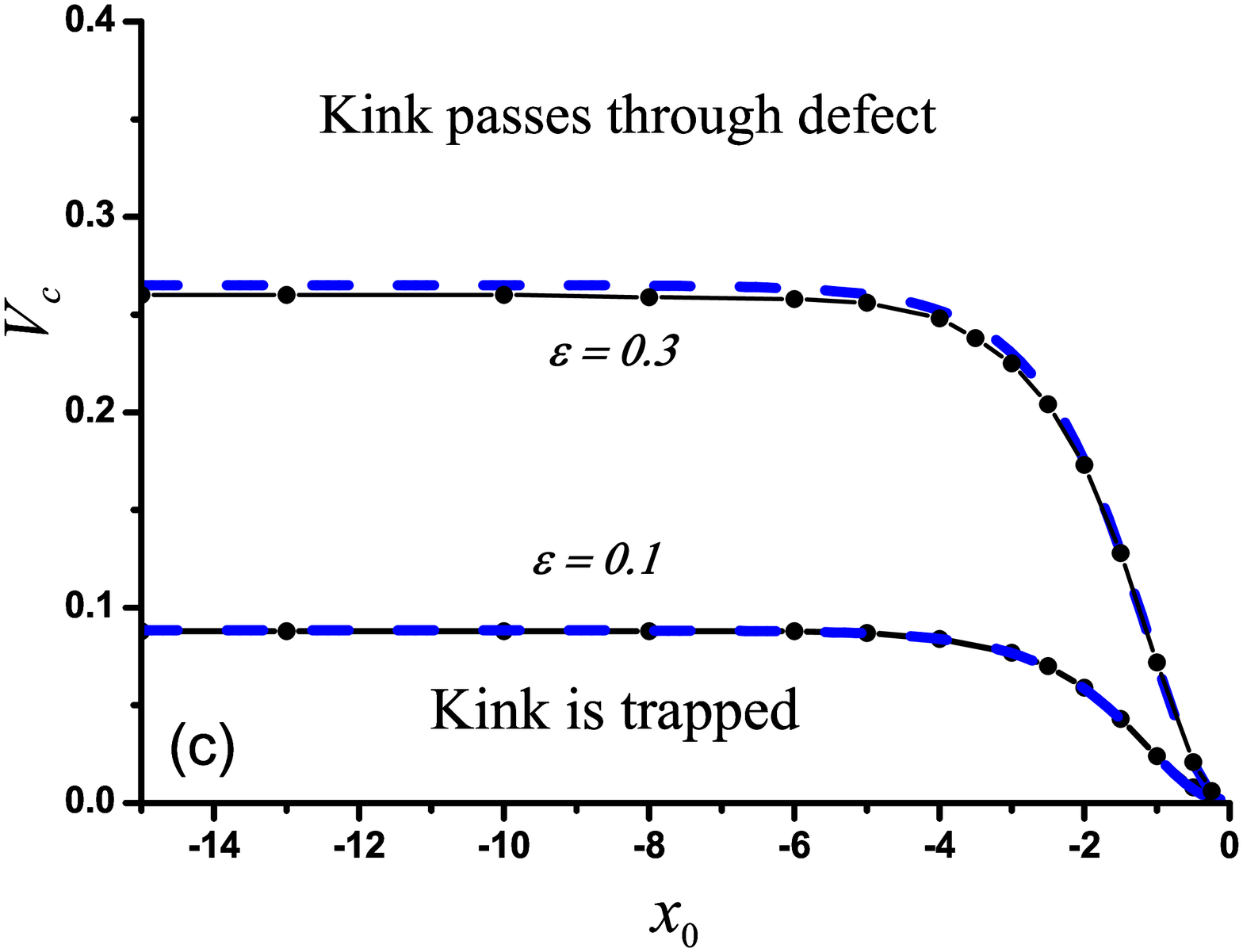}
\caption{(Color online) (a,b) Kink position as a function of time
for the initial velocity $V_k=\pm0.1$, for the case when the kink
approaches the defect (a) from the gain side, (b) from the loss
side. Defect center is located at $x=0$. (c) Kink critical
velocity as a function of its initial position. The results for
the perturbation amplitudes $\epsilon=0.1$ and $\epsilon=0.3$ are
presented. Solid lines show the results of the numerical solution for
the continuous system and dash lines show the results obtained
with the help of the collective variable method.} \label{fig1}
\end{figure}

In Fig.~\ref{fig1}~(a) the kink position as a function of time for
the kink moving toward the defect from the gain side with the
initial velocity $V_k=-0.1$ is presented for the defect amplitudes
$\epsilon=0.1$ and $\epsilon=0.3$, as indicated for each curve.
Solid lines present the result of numerical integration of
Eq.~(\ref{FK_Collisions}), while dashed lines give the result of
numerical integration of the collective variable model
Eq.~(\ref{Coll_var1}). One can see that the collective variable
approach gives a very accurate prediction of the actual kink
dynamics. From this figure it is clearly seen that the kink moving
toward the defect from the gain side is first accelerated and
after passing the gain side of the defect it is decelerated by the
loss side. After the kink passes the defect and moves far from it,
it seems to restore its initial velocity and energy. The effects
of the kink-defect interaction in this case are the phase shift
and also the excitation of the kink's internal mode which will be
discussed in the following.

Next, suppose a kink comes from the lossy side of the defect. In
this case, two different scenarios for the kink interaction with
the defect are possible depending on its initial velocity $V_k$ or
defect strength $\epsilon$, as presented in Fig.~\ref{fig1}~(b)
for $\epsilon=0.1$ and $\epsilon=0.3$ with $V_k=0.1$ in both
cases. If $V_k$ is large enough or $\epsilon$ is small enough, the
kink passes through the lossy part of the defect with the velocity
smaller than $V_k$ and enters the gain part where it is
accelerated up to the initial velocity and then goes on to
infinity. In the opposite case (where $V_k$ is not large enough or
$\epsilon$ is not small enough), the kink does not possess
sufficient momentum to pass through the lossy part of the defect
and it is trapped there. As one can see from Fig.~\ref{fig1}~(b),
for the case of $\epsilon=0.1$, the kink passes through the defect and
effectively restores its initial velocity, while for
$\epsilon=0.3$ the kink is trapped by the lossy side of the
defect.

In Fig.~\ref{fig1}~(c) the kink critical velocity is shown as the
function of kink initial position for the two values of defect
strength, $\epsilon=0.1$ and $\epsilon=0.3$. Numerical results
(solid lines) are in a very good agreement with the collective
variable results (dashed lines) given by Eq.~(\ref{Coll_var5}) for
the case of $\beta=\delta_k=1$. The results suggest that for the
kink initially located within the loss region of the defect the
value of the critical velocity is smaller the smaller $\epsilon$
is (which is intuitively clear), as well as smaller the closer one
starts to the $x_0=0$, i.e., to the center of the
$\mathcal{PT}$-symmetric defect.

\begin{figure}
\includegraphics[width=9.5cm]{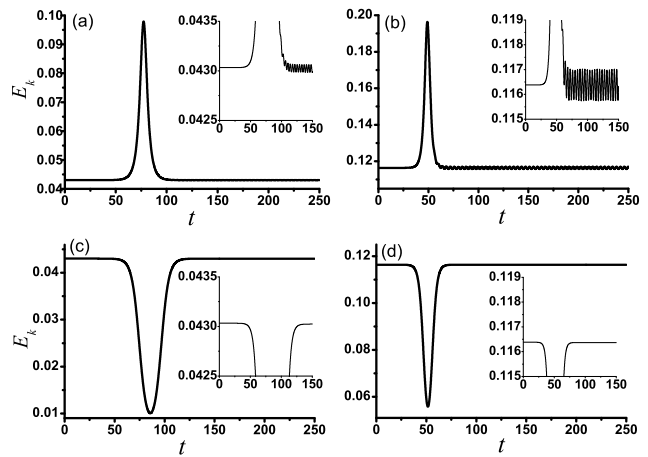}
\caption{Numerical results for the continuum model showing the
interaction of the kink initially bearing no IM with the
$\mathcal{PT}$-symmetric defect of strength $\epsilon=0.15$. The
kink approaches the defect from (a,b) the gain side and (c,d) from
the lossy side of the defect. Shown is the kink kinetic energy as
a function of time. The kink initial velocity is (a,c) $V_k=0.25$
and (b,d) $V_k=0.4$. Different scale used for insets reveals that
in (a,b) kink's IM is excited after the interaction with the
defect, while in (c,d) it is not excited.} \label{fig2}
\end{figure}

Fig.~\ref{fig2} shows the kink kinetic energy as a function of
time for the case when the kink, initially bearing no IM,
interacts with the $\mathcal{PT}$-symmetric defect with amplitude
$\epsilon=0.15$. In (a,b) the kink approaches the defect from the
gain side and in (c,d) from the loss side. The kink initial
velocity is $V_k=0.25$ in (a,c) and $V_k=0.4$ in (b,d). As a
result of interaction with the defect, the kink in (a,b) is
firstly accelerated and then decelerated, while in (c,d) it is
first decelerated and then accelerated. In both cases the kink's
translational velocity after passing through the defect is
practically identical to the initial velocity. Note that the kink
without the IM excited has a constant in time kinetic energy,
while the kinetic energy of the kink with the excited IM
oscillates near the constant value with frequency $2\sqrt{3}$,
which is double the IM frequency. The insets in (a,b) show that
when the kink hits the defect from the gain side, a noticeable IM
is excited as a result of the kink-defect interaction, while for
the kink coming from the opposite direction [see insets in (c,d)],
the IM is much weaker and, in fact, cannot be seen in the scale of
the figure. The effect of kink's IM excitation as a result of the
interaction with the defect becomes stronger for larger initial
kink velocity $V_k$, as can be seen from comparison of the insets
in (a) and (b). However, it is clear that the presence of the IM
does not preclude the kink from traveling through the defect, when
it comes from the gain side. In fact, as we will see now, quite
the opposite is true.

\begin{figure}
\includegraphics[width=8.5cm]{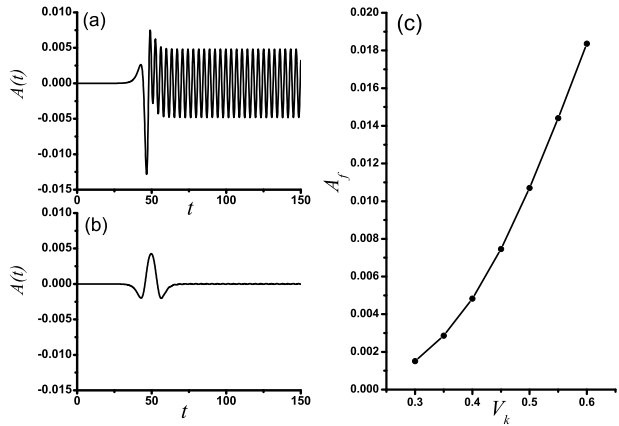}
\caption{Collective variable results for the model
Eqs.~(\ref{Coll_var1a},\ref{Coll_var1aa}). (a,b) The dynamics of
the internal shape mode whose amplitude $A(t)$ is shown for the
case when the kink comes from (a) gain side and (b) loss side. The
perturbation amplitude is $\epsilon=0.15$ and the the initial
velocity of the kink is $V_k=\pm0.4$. (c) Amplitude of the shape
mode after the interaction with the defect as a function of kink
initial velocity for the case when kink hits the defect from the
gain side.} \label{fig3}
\end{figure}

The results obtained for the partial differential equation of
(\ref{phi4}) and presented in Fig.~\ref{fig2} will be now compared
to the results of the numerical solution of Eqs.
(\ref{Coll_var1a},\ref{Coll_var1aa}) for the two degree of freedom
collective variable model, see Fig.~\ref{fig3}. In (a,b) we plot
the amplitude of the shape mode as the function of time. Here the
kink with initial velocity $V_k=\pm0.4$ hits the defect of
amplitude $\epsilon=0.15$ from the gain side (Fig.~\ref{fig3}~(a))
and loss side (Fig.~\ref{fig3}~(b)). It can be seen that for the
kink coming from the gain side a noticeable IM is excited after
the interaction with the defect, whereas for the kink moving in
the opposite direction the residual excitation is much weaker and
is barely visible in the scale of the figure. This is in very good
qualitative agreement with the results for continuum model. In
Fig.~\ref{fig3}~(c) the amplitude of the shape mode after the
interaction with the defect is plotted as the function of kink
initial velocity for the case when kink hits the defect from the
gain side. The plot (see Fig.~\ref{fig3}~(c)) shows that the
amplitude increases with increasing kink initial velocity. Again,
this is in line with the observations made for the continuum
system.

\subsection{Kinks with initially excited IM}

Here, the scattering of a kink bearing an IM on $\mathcal{PT}$-symmetric
defect is considered. The intensity of the IM will be characterized by
the amplitude of the kink kinetic energy oscillation caused by the
IM, $\Delta E_k=(E_{k,\max}-E_{k,\min})/2$, where $E_{k,\max}$ and
$E_{k,\min}$ are the maximum and minimum of the kink kinetic
energy.

\begin{figure}
\includegraphics[width=8.5cm]{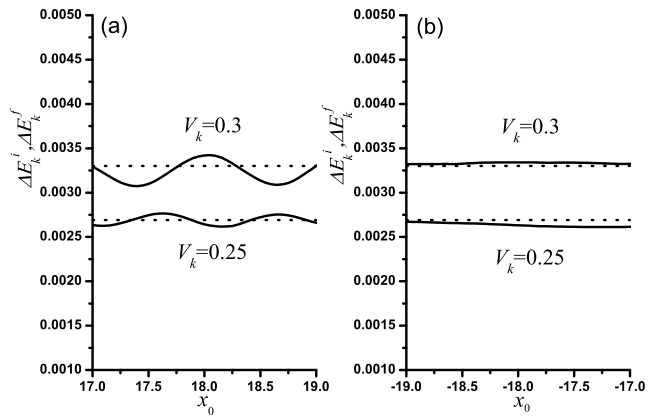}
\caption{Numerical results for the continuum model. Amplitude of
the kink kinetic energy oscillation before ($\Delta E_k^i$, dotted
lines) and after ($\Delta E_k^f$, solid lines) the interaction
with the $\mathcal{PT}$-symmetric defect as a function of kink
initial position for the two different values of the kink initial
velocities, $V_k=0.25$ and $V_k=0.3$. The defect amplitude is
$\epsilon=0.15$. In (a) the kink comes from the gain side and (b) from
the loss side.} \label{fig4}
\end{figure}

The kink with the IM is excited with the help of
Eq.~(\ref{KinkModeSolution}) using the IM amplitude $A=0.05$ and
two values of initial kink velocity, $V_k=0.25$ and $V_k=0.3$. For
these two cases the amplitude of the initial kink kinetic energy
oscillation, $\Delta E_k^i$, is equal to 0.00269 and 0.00330,
respectively. These initial values are plotted in Fig.~\ref{fig4}
with horizontal dotted lines. Now we calculate the value of
$\Delta E_k^f$ after the kink passes through the defect with the
amplitude $\epsilon=0.15$ as a function of the kink initial
position and present the results by solid lines in Fig.~\ref{fig4}
for the kink moving from (a) the gain and (b) the loss side of the
defect. In (a), it is clear that the kink's IM is affected by the
defect because $\Delta E_k^f$ differs from $\Delta E_k^i$, while
in (b) they almost coincide, which means that the IM amplitude is
not changed by the defect. The oscillation of $\Delta E_k^f$ as
the function of kink initial position $x_0$ in (a) has the period
close to $V_k(2\pi/\omega)$, which is the distance the kink
travels in one period of IM oscillation.

\begin{figure}
\includegraphics[width=8.5cm]{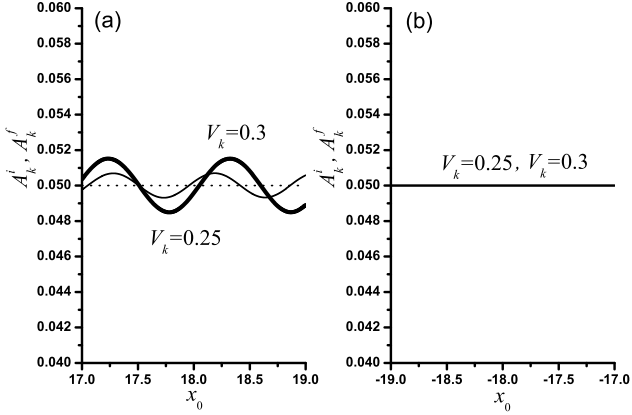}
\caption{Collective variable results for the model
Eqs.~(\ref{Coll_var1a},\ref{Coll_var1aa}). The amplitude of the
kink's internal shape mode before ($A_k^i$, dotted lines) and
after ($A_k^f$, solid lines) the interaction with the
$\mathcal{PT}$-symmetric defect as a function of the kink initial
position for the two different values of the kink initial
velocities, $V_k=0.25$ (thin line) and $V_k=0.3$ (thick line). The
defect amplitude is $\epsilon=0.15$. In (a) the kink comes from
the gain side and (b) from the loss side.} \label{fig5}
\end{figure}

Similar results obtained in frame of the two degree of freedom
collective variable model
Eqs.~(\ref{Coll_var1a},\ref{Coll_var1aa}) are presented in
Fig.~\ref{fig5}. The initial ($A_k^i$) and final ($A_k^f$)
amplitudes of the kink's shape mode are shown as functions of the
kink initial position by the dotted and solid lines, respectively.
In (a) the kink hits the defect with amplitude $\epsilon=0.15$
from the gain side and in (b) from the loss side. The kink
initially has IM with the amplitude $A=0.05$ and the initial
velocities of the kinks are $V_k=0.25$ (thin line) and $V_k=0.3$
(thick line) in both cases. In very good qualitative agreement
with the results for continuum model, the collective variable
model shows that in (a) the kink's IM is affected by the defect,
while in (b) it is not, since all the three lines overlap.

\begin{figure}
\includegraphics[width=8.5cm]{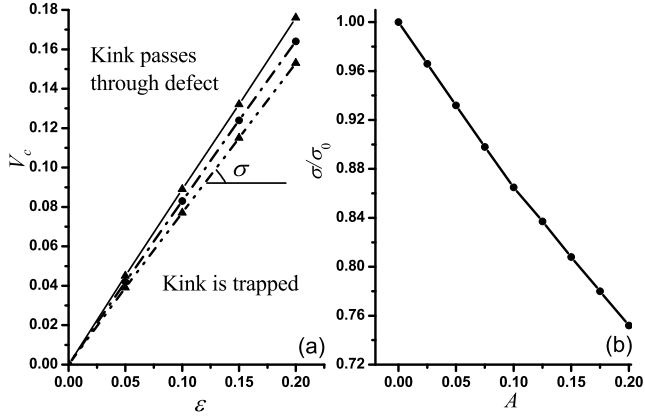}
\caption{Results for the continuum model. (a) Relation between the
critical initial velocity of the kink and the defect amplitude for
the kink with IM amplitude $A=0$ (solid line), $A=0.05$ (dot-dash
line), and $A=0.1$ (dot-dot-dash line). The dependencies are
nearly linear with the slope $\sigma$ decreasing with an increase
in the IM amplitude $A$. (b) The value $\sigma/\sigma_0$ as the
function of $A$, where $\sigma_0$ corresponds to $A=0$.}
\label{fig6}
\end{figure}

The results shown in Fig.~\ref{fig6} demonstrate that the kink
bearing an excited IM penetrates more easily through the
dissipative media. In Fig.~\ref{fig6}~(a), the kink critical
velocity as a function of the perturbation strength $\epsilon$ is
shown for the kink bearing an IM with the amplitude $A=0$ (solid
line), $A=0.05$ (dot-dashed line), and $A=0.1$ (dot-dot-dashed
line). One can see that $V_c$ is almost linear function of
$\epsilon$ with a slope denoted by $\sigma$. It is clear that for
given perturbation strength $\epsilon$ the critical velocity
decreases with increase in the IM amplitude $A$. In
Fig.~\ref{fig6}~(b) we plot the value of $\sigma/\sigma_0$ as a
function of $A$, where $\sigma_0$ corresponds to $A=0$. The slope
$\sigma$ decreases with an increasing IM amplitude indicating that
the translational degree of freedom of the kink when the IM is
excited is less affected by the perturbation. That is to say, the
transparency of the defect to the kink transmission is increased
as the IM excitation is increased.

\begin{figure}
\includegraphics[width=8.5cm]{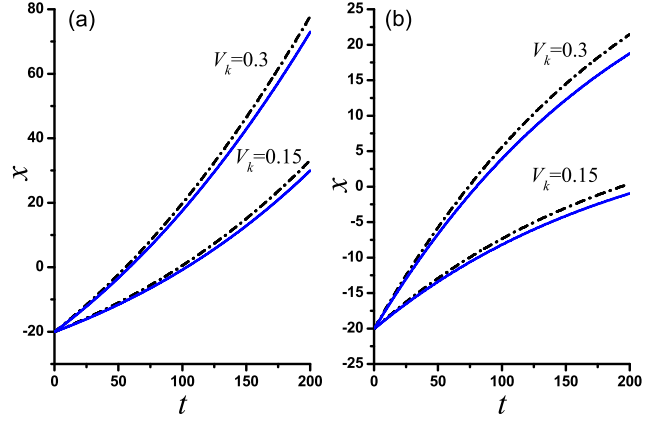}
\caption{(Color online) Results for the continuum model. Dynamics
of kinks with initial velocities equal to 0.15 and 0.3 in the case
of homogeneous perturbation $\gamma(x)\equiv1$ with (a)
$\epsilon=0.005$ (gain) and (b) $\epsilon=-0.005$ (loss). Solid
lines are for the IM amplitude $A=0$ and dash-dot lines for
$A=0.05$. The kink trajectories show that the velocity of the kink
gradually increases in (a) (decreases in (b)), and the kink
velocity is always higher for the kink with IM.} \label{fig7}
\end{figure}

\begin{figure}
\includegraphics[width=8.5cm]{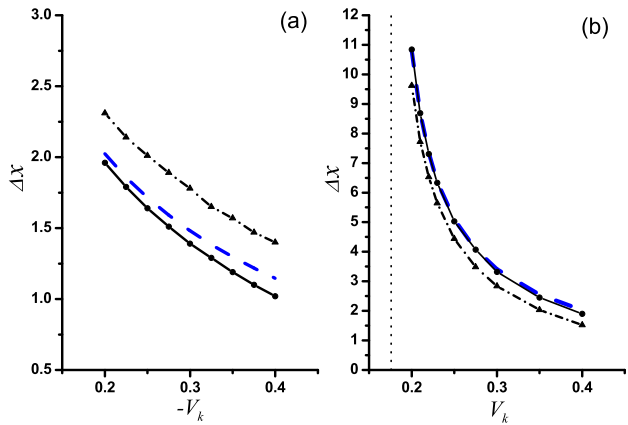}
\caption{(Color online) Results for the continuum model and
collective variable results of Eqs.~(\ref{Denis19},\ref{Denis23}).
The kink's phase shift due to the interaction with the
$\mathcal{PT}$-symmetric defect as a function of kink initial
velocity for the kink moving toward the defect with strength
$\epsilon=0.2$ (a) from the gain side and (b) from the loss side.
Solid lines show the results of numerical solution for the kink
without IM, dot-dash lines for the kink bearing IM with amplitude
$A=0.01$. Dashed lines show the results obtained with the help of
the collective variable method. The vertical dotted line in (b)
shows the threshold kink velocity $V_c=0.176$.} \label{fig8}
\end{figure}

In Fig.~\ref{fig7} we contrast the dynamics of the kinks with and
without IM in the case of homogeneous perturbation
$\gamma(x)\equiv 1$ with (a) $\epsilon=0.005$  (purely gain, hence
acceleration) and (b) $\epsilon=-0.005$  (purely loss, hence
deceleration). Two values of initial kink velocities are
considered in both cases, namely, 0.15 and 0.3. The kink center
position as a function of time is shown for the cases of no IM,
$A=0$ (solid lines), and for the IM of amplitude $A=0.05$
(dash-dotted lines). It can be seen that in all cases the kink
with an excited IM travels faster than the kink without IM. In
other words, the kink bearing IM in the gain media is accelerated
faster and in the loss media it is decelerated slower. From here
it immediately follows that the phase shift due to interaction
with the $\mathcal{PT}$-symmetric defect must be higher (lower)
for the kink with IM when it comes from the gain (loss) side. This
is illustrated by Fig.~\ref{fig8} in which the kink's phase shift
due to interaction with the $\mathcal{PT}$-symmetric defect of
strength $\epsilon=0.2$ is presented as a function of its initial
velocity. Solid lines show the results of the numerical solution
for the kink without IM, dot-dashed lines indicate the kink
bearing an IM with amplitude $A=0.01$. Note that in (a) the kink
moves toward the defect from the gain side and the phase shift is
higher for the kink bearing IM. In (b) the kink approaches the
defect from the opposite side and the phase shift is higher for
the kink without IM. The vertical dotted line in (b) shows the
critical value of the initial kink velocity, $V_c$. For
comparison, dashed lines in Fig.~\ref{fig8} show the results
obtained with the help of the collective variable method of
Eqs.~(\ref{Denis19}) and (\ref{Denis23}). Note that the accuracy
of the collective variable method is very high for the kink
without IM.

\section {Conclusions} \label{Sec:V}

Interaction of the $\phi^4$ kinks with and without IM excitation
with the $\mathcal{PT}$-symmetric defect having balanced regions
of positive and negative dissipation was investigated numerically,
as well as analytically with the help of the collective variable
model.

Similar to the SG kink studied earlier \cite{Danial}, a $\phi^4$
kink coming from the gain side always passes through the defect
and practically restores its initial velocity (see
Fig.~\ref{fig1}~(a)). For the kink approaching the defect from the
opposite side, there exist two different scenarios, depending on
the kink initial velocity $V_k$. For $V_k<V_c$, where $V_c$ is a
threshold value of the velocity, the kink does not have enough
energy to pass through the defect and it is trapped by the lossy
side of the defect (see Fig.~\ref{fig1}~(b)), while for $V_k>V_c$
it enters the gain region and effectively restores initial
velocity.

In contrast to the SG kink, the $\phi^4$ kink possesses an
internal mode. From our numerical results presented in
Fig.~\ref{fig2} it follows that a noticeable IM is excited on the
$\phi^4$ kink after passing the defect from the gain side and the
excitation of IM is much weaker for the kink moving in the
opposite direction. The excitation of the kink's IM increases with
increasing kink initial velocity; see Fig.~\ref{fig2} (a) and (b).
These effects are well reproduced by the two degree of freedom
collective variable model
Eqs.~(\ref{Coll_var1a},\ref{Coll_var1aa}), as shown in
Fig.~\ref{fig3}.

On the other hand, a kink moving from the gain side with an
initially excited IM passes through the defect with the IM having,
generally speaking, different amplitude, and the effect is
stronger for faster kinks [see Fig.~\ref{fig4} (a)]. For the kink
moving in the opposite direction the initially excited IM is not
affected by the defect [see Fig.~\ref{fig4} (b)]. Analogous
results obtained with the use of the collective variable model
(see Fig.~\ref{fig5}) are in good agreement with the results
for the continuum $\phi^4$ model shown in Fig.~\ref{fig4}.

The kink bearing an IM is faster accelerated in the gain region
and slower decelerated in the loss region of the defect in
comparison to the kink free of the initially excited IM. As a
result, (i) the critical velocity to pass through the loss region
is smaller for the kink with IM (see Fig.~\ref{fig6}) and (ii) the
phase shift due to the passage through the defect increases when
the kink with IM hits the defect from the gain side and decreases
in the opposite case [see Fig.~\ref{fig8} (a) and (b)].

The single degree of freedom collective variable model gives a
very good prediction of the dynamics of the kink's center.
Analytical expressions for the kink phase shift and kink critical
velocity were derived in frame of this model.

We conclude that the $\mathcal{PT}$-symmetric defects give new
opportunities in the manipulation of the soliton dynamics
and the presence of the internal modes can induce noticeable
asymmetries of the solitary wave-defect interaction
both in terms of their
excitation, as well as of the kink transmission in the presence
of such a mode.

In future work, it would be of interest to study the $\phi^4$ kink
dynamics in the case of $\mathcal{PT}$-symmetric, periodic
$\gamma(x)$. According to the results presented here, kink
acceleration can be expected when an IM is excited, even though
the gain and loss is balanced in the model. Consideration of the
fields with radial symmetry~\cite{jeanguy} and models of higher
dimension is also very tempting. In the one dimensional realm,
however, it would also be of particular interest to explore the
very rich setting of kink-antikink collisions in the presence of
defects. Finally, while a detailed understanding is at this point
available based on collective coordinate techniques in the
Klein-Gordon field theoretic setting, the picture of
soliton-defect interaction and the role of internal modes is far
less clear in nonlinear Schr{\"o}dinger equation settings, where
such questions are just starting to be explored~\cite{natanael}.
Hence, it would be of particular value to expand on such studies
in the near future. These topics are presently under investigation
and will be reported in future publications.

\section*{Acknowledgments}

D.S. thanks the hospitality of the Bashkir State Pedagogical
University and acknowledges the financial support from the
Institute for Metals Superplasticity Problems, Ufa, Russia. S.V.D.
thanks financial support provided by the Russian Government
Program 5-100-2020. D.I.B. was partially supported by a grant of
Russian Foundation for Basic Research, the grant of the President
of Russian Federation for young scientists-doctors of science
(project no. MD-183.2014.1) and by the fellowship of Dynasty
foundation for young Russian mathematicians. P.G.K. acknowledges
support from the US National Science Foundation under grants
CMMI-1000337, DMS-1312856, from the ERC and FP7-People under grant
IRSES-606096 from the Binational (US-Israel) Science Foundation
through grant 2010239, and from the US-AFOSR under grant
FA9550-12-10332.

\section{Appendix}
\label{Appendix}

An attempt to apply commercial symbolic derivation software did
not generically provide a satisfactory result. For this reason, in
Sec.~\ref{AppendixA} we calculate analytically the kink critical
velocity to pass through the lossy part of the defect and the kink
phase shift due to the interaction with the defect. In
Sec.~\ref{AppendixB}, for the particular case of $\be=\dl$, the
phase shift is given in a closed form.

\subsection{Analytic results}
\label{AppendixA}

Given any two numbers $\cX_0$ and $\cX$, we denote
\begin{equation}\label{eq1}
\begin{aligned}
g(\cX_0,\cX):=&\int\limits_{-\infty}^{+\infty} dx
\int\limits_{\cX_0}^{\cX} \frac{\g(x) ds}{\cosh^4(\dl(x-s))}
\\
=& \int\limits_{-\infty}^{+\infty} dx \,
\g(x)\int\limits_{\cX_0}^{\cX} \frac{ ds}{\cosh^4 \dl(x-s)}.
\end{aligned}
\end{equation}
It is clear that for our choice of $\gamma(x)$
\begin{equation}\label{eq2}
\g(x)=-\frac{1}{\be} \frac{d\hphantom{x}}{dx} \frac{1}{\cosh \be
x}
\end{equation}
and thus we can integrate by parts in (\ref{eq1}):
\begin{align*}
g(\cX_0,\cX)=\frac{1}{\be} \int\limits_{-\infty}^{+\infty} &
\frac{1}{\cosh\be x}\left(\frac{1}{\cosh^4(\dl(x-\cX))}\right.
\\
&\left.- \frac{1}{\cosh^4\dl(x-\cX_0)}\right) dx.
\end{align*}
We introduce the notation
\begin{equation}\label{eq3}
G(\cX):=\frac{1}{\be}\int\limits_{-\infty}^{+\infty}
\frac{dx}{\cosh\be x \cosh^4\dl(x-\cX)}
\end{equation}
and see that
\begin{equation}\label{eq4}
g(\cX_0,\cX)=G(\cX)-G(\cX_0).
\end{equation}
By a simple change of variable $x\mapsto\dl x$ we reduce the
formula for $G(\cX)$ to
\begin{equation}\label{eq5}
G(\cX)=\frac{1}{\be\dl} \int\limits_{-\infty}^{+\infty}
\frac{dx}{\cosh \frac{\be}{\dl} x \cosh^4 (x-\dl \cX)}.
\end{equation}
Now we make our main assumption:
\begin{equation}\label{eq6}
\be=\frac{2m-1}{n}\dl,\quad n,m\in\mathds{N}.
\end{equation}
This assumption is quite general in the sense that given any $\be$
and $\dl$, we can always approximate with any prescribed accuracy
the fraction $\be/\dl$ choosing appropriate $m$ and $n$.

In view of (\ref{eq6}) formula (\ref{eq5}) casts into the form
\begin{equation*}
G(\cX)=\frac{n}{(2m-1)\dl^2}  \int\limits_{-\infty}^{+\infty}
\frac{dx}{\cosh \frac{2m-1}{n} x \cosh^4 (x-\dl \cX)}.
\end{equation*}
Here we make one more change of variable $x=n\ln y$. It yields
\begin{equation}\label{eq7}
G(\cX)=\frac{2^5 n^2}{(2m-1)\dl^2} \E^{4\dl \cX} G_1(\E^{2\dl
\cX}),
\end{equation}
where
\begin{align*}
G_1(\cY):=&\int\limits_{0}^{+\infty}
\frac{y^{4n+2m-2}dy}{(y^{4m-2}+1)(y^{2n}+\cY)^4}
\\
=&\frac{1}{2}\int\limits_{-\infty}^{+\infty}
\frac{y^{4n+2m-2}dy}{(y^{4m-2}+1)(y^{2n}+\cY)^4},\quad \cY>0.
\end{align*}
It is straightforward to check that
\begin{equation}\label{eq8}
\begin{aligned}
&G_1(\cY)=\frac{1}{12} \frac{d^2 G_2}{d\cY^2}(\cY),
\\
&G_2(\cY):=\int\limits_{-\infty}^{+\infty}
\frac{y^{4n+2m-2}dy}{(y^{4m-2}+1)(y^{2n}+\cY)^2}.
\end{aligned}
\end{equation}
The differentiation w.r.t. a parameter $\cY$ is possible, since
the integral in the definition of $G_2(\cY)$ converges uniformly
in $\cY$. We also observe that although formally it is possible,
we do not pass to the third derivative w.r.t. $\cY$ in the latter
formula since it leads us to a diverging integral.

We calculate function $G_2$ by means of complex analysis. Namely,
we employ an approach based on finding residues, see for instance,
\cite[Ch. V\!I, Sect. 3]{Evgrafov}. In accordance with these
approach, we first regard $y$ as a complex variable and consider
the integrand in (\ref{eq8}) as a meromorphic function defined on
the complex plane $\mathds{C}$. Then in accordance with the Cauchy
theorem function $G_2$ is given by the formula:
\begin{equation}\label{eq9}
G_2(\cY)=2\pi\iu \sum\limits_{j} \res\limits_{y=y_j}
\frac{y^{4n+2m-2}}{(y^{4m-2}+1)(y^{2n}+\cY)^2},
\end{equation}
where $\iu$ stands for the imaginary unit and the sum is taken
over all the poles in the complex upper half-plane $\IM y>0$. The
symbol $\res\limits_{y=y_j} f$ denotes the residue of the function
at a pole $y_j$. We recall that the residue at $y=y_j$ of a
meromorphic function is a coefficient at the power $(y-y_j)^{-1}$
in the Laurent expansion for a function at point $y_j$.

In our case the poles are exactly the roots of the equation
\begin{equation*}
(y^{4m-2}+1)(y^{2n}+\cY)^2=0.
\end{equation*}
The roots located in the complex upper half-plane are
\begin{align*}
&y_j^{(m)}=\E^{\frac{\iu\pi(2j+1)}{4m-2}},\quad j=0,\ldots,2m-2,
\\
&y_j^{(n)}=\cY^{\frac{1}{2n}}\E^{\frac{\iu\pi(2j+1)}{2n}},\quad
j=0,\ldots,n-1.
\end{align*}
To calculate the residues at these poles, we need to find the
coefficients $c_j^{(\natural)}$ in the expansions
\begin{equation}\label{eq9c}
\frac{y^{4n+2m-2}}{(y^{4m-2}+1)(y^{2n}+\cY)^2}=\ldots
+\frac{c_j^{(\natural)}}{y-y_j^{(\natural)}}+\ldots,
\end{equation}
as $y\to y_j^{(\natural)}$, where $\natural=n,m$.

Suppose $\cY\not=1$ and let us find first the desired coefficients
for $y_j^{(m)}$. We should just represent $(y^{4m-2}+1)$ as
\begin{equation}\label{eq9a}
y^{4m-2}+1=(y-y_j^{(m)})P_j^{(m)}(y),
\end{equation}
where $P_j^{(m)}$ is a some polynomial. Then
\begin{equation*}
c_j^{(m)}=\frac{(y_j^{(m)})^{4n+2m-2}}{P_j^{(m)}(y_j^{(m)})
\left((y_j^{(m)})^{2n} +\cY\right)^2}.
\end{equation*}
Differentiating (\ref{eq9}) w.r.t. $y$ and letting then
$y=y_j^{(m)}$, we get
\begin{equation}\label{eq9b}
P_j^{(m)}(y_j^{(m)})=(4m-2)(y_j^{(m)})^{4m-3}=-\frac{4m-2}{y_j^{(m)}}.
\end{equation}
Hence,
\begin{equation}\label{eq10}
\begin{aligned}
c_j^{(m)}=&-\frac{(y_j^{(m)})^{2m+4n-1}}{(4m-2)\left((y_j^{(m)})^{2n}+\cY\right)^2}
\\
=&\frac{(y_j^{(m)})^{4n-1}}{(4m-2)\left((y_j^{(m)})^{2n}+\cY\right)^2}
\\
=&\frac{\E^{\frac{\iu\pi}{4m-2}(2j+1)(4n-1)}}{(4m-2)\left(\E^{\frac{\iu\pi
n}{2m-1}(2j+1)}+\cY\right)^2},
\end{aligned}
\end{equation}
where $j=0,\ldots,2m-2$.

The calculations of the residues for $y_j^{(n)}$ are more tricky
since the function in the right hand side of (\ref{eq8}) has
second order poles at $y_j^{(n)}$. By analogy with (\ref{eq9a}) we
introduce the representation
\begin{equation}\label{eq11}
y^{2n}+\cY=(y-y_j^{(n)})P_j^{(n)}(y)
\end{equation}
and differentiating it once and twice and letting then
$y=y_j^{(n)}$, in the same way as in (\ref{eq9b}) we obtain
\begin{equation}\label{eq12}
\begin{aligned}
&P_j^{(n)}(y_j^{(n)})=-\frac{2n\cY}{y_j^{(n)}},
\\
&\frac{d
P_j^{(n)}}{dy}(y_j^{(n)})=-\frac{n(2n-1)\cY}{(y_j^{(n)})^2}.
\end{aligned}
\end{equation}
Let $h(y):=y^{4n +2m-2}/(y^{4m-2}+1)$, then it follows from
(\ref{eq9c}) that
\begin{align*}
c_j^{(n)}=&\frac{d\hphantom{y}}{dy}
\frac{h}{(P_j^{(n)})^2}\bigg|_{y=y_j^{(n)}}
\\
=&\frac{h'(y_j^n)}{\big(P_j^{(n)}(y_j^{n})\big)^2}
-\frac{2h(y_j^{(n)}){P_j^{(n)}}'(y_j^{(n)})}{\big(P_j^{(n)}(y_j^{(n)})\big)^3}.
\end{align*}
Thus, due to the definition of $h$, $y_j^{(n)}$, and (\ref{eq12})
\begin{align*}
c_j^{(n)}=&\frac{h'(y_j^n) (y_j^{(n)})^2-(2n-1)  h(y_j^{(n)})
y_j^{(n)}}{4n^2 \cY^2}
\\
=&\frac{h(y_j^{(n)})
y_j^{(n)}}{4n^2\cY^2}\left(y_j^{(n)}\frac{d\ln h}{dy}(y_j^n)
-2n+1\right)
\\
=&\frac{1}{4n^2}
\left(2n+2m-1+\frac{4m-2}{(y_j^{(n)})^{4m-2}+1}\right)
\\
&\cdot\frac{(y_j^{(n)})^{2m-1}}{(y_j^{(n)})^{4m-2}+1}
\\
=&\frac{1}{4n^2}
\left(2n-2m-1+\frac{4m-2}{\cY^{\frac{2m-1}{n}}\E^{\frac{\iu\pi(2j+1)(2m-1)}{2n}}+1}\right)
\\
&\cdot
\frac{\cY^{\frac{2m-1}{2n}}\E^{\frac{\iu\pi(2j+1)(2m-1)}{2n}}}{\cY^{\frac{2m-1}{n}}
\E^{\frac{\iu\pi(2j+1)(2m-1)}{n}}+1}.
\end{align*}
We substitute the above formulae and (\ref{eq10}) for the residues
into (\ref{eq9}) and obtain the final expression for $G_2$:
\begin{align*}
G_2&(\cY)=\frac{\pi\iu}{2m-1} \sum\limits_{j=0}^{2m-2}
\frac{\E^{\frac{\iu\pi(2j+1)(4n-1)}{4m-2}}}{\left(\E^{\frac{\iu\pi
n}{4m-2}(2j+1)}+\cY\right)^2}
\\
&+\frac{\pi\iu}{2n^2} \sum\limits_{j=0}^{n-1}
\left(2n-2m-1+\frac{4m-2}{\cY^{\frac{2m-1}{n}}\E^{\frac{\iu\pi(2j+1)(2m-1)}{2n}}+1}\right)
\\
&\hphantom{+\frac{\pi\iu}{2n^2}
\sum\limits_{j=0}^{n-1}\Bigg(}\cdot
\frac{\cY^{\frac{2m-1}{2n}}\E^{\frac{\iu\pi(2j+1)(2m-1)}{2n}}}{\cY^{\frac{2m-1}{n}}
\E^{\frac{\iu\pi(2j+1)(2m-1)}{n}}+1}.
\end{align*}
In view of (\ref{eq8}) it leads us to the formula for $G_1$:
\begin{equation}\label{eq13}
\begin{aligned}
&G_1(\cY)=\frac{\pi\iu}{4m-2} \sum\limits_{j=0}^{2m-2}
\frac{\E^{\frac{\iu\pi(2j+1)(4n-1)}{4m-2}}}{\left(\E^{\frac{\iu\pi
n}{2m-1}(2j+1)}+\cY\right)^4}
\\
&+\frac{\pi\iu}{24n^2} \sum\limits_{j=0}^{n-1}
\frac{d^2\hphantom{\cY}}{d\cY^2}
\left(2n-2m-1+\frac{4m-2}{\cY^{\frac{2m-1}{n}}\E^{\frac{\iu\pi(2j+1)(2m-1)}{2n}}+1}\right)
\\
&\hphantom{+\frac{\pi\iu}{2n^2}
\cY^{\frac{2m-1}{2n}}\sum\limits_{j=0}^{n-1}\Bigg(}\cdot
\frac{\cY^{\frac{2m-1}{2n}}\E^{\frac{\iu\pi(2j+1)(2m-1)}{2n}}}{\cY^{\frac{2m-1}{n}}
\E^{\frac{\iu\pi(2j+1)(2m-1)}{n}}+1}.
\end{aligned}
\end{equation}

Although in the above calculations we have supposed that
$\cY\not=1$, it is clear that $G_1$ is a continuous function and
thus we can find $G_1(1)$ as
\begin{equation*}
G_1(1)=\lim\limits_{\cY\to 1} G_1(\cY).
\end{equation*}
It also follows from (\ref{eq4}), (\ref{eq7}) that the critical
velocity is given by the formula
\begin{equation}\label{eq14}
\begin{aligned}
V_c=&\frac{\epsilon\dl^2}{M} g(0,\cX_0)
\\
=&\frac{32n^2 \epsilon\dl^2}{M(2m-1)}\left( G_1(1)-\E^{4\dl
\cX_0}G_1(\E^{2\dl \cX_0})\right),
\end{aligned}
\end{equation}
where $G_1$ is determined by the above formulae.

In particular, if $n=m=1$, then $\be=\dl$,
\begin{align}
&
\begin{aligned}
G_1(\cY)&=\frac{\pi}{32} \frac{\cY^3-9\cY^2+
16\cY^{\frac{3}{2}}-9\cY+1}{\cY^{\frac{3}{2}}(\cY-1)^4}
\\
&=\frac{\pi}{32}
\frac{\cY+4\cY^{\frac{1}{2}}+1}{\cY^{\frac{3}{2}}(\cY^{\frac{1}{2}}+1)^4},
\end{aligned}\label{eq14a}
\\
&G_1(1)=\frac{3\pi}{256}.\nonumber
\end{align}
For $n=2$, $m=1$ we have $\be=\dl/2$,
\begin{equation}\label{eq14aa}
\begin{aligned}
&G_1(\cY)=\frac{\pi}{512\cY^{\frac{7}{4}}(1+\cY)^4}\big(5\sqrt{2}\cY^{\frac{7}{2}}
-7\sqrt{2}\cY^3+35\sqrt{2}\cY^{\frac{5}{2}}
\\
&-105\sqrt{2}\cY^2 -105\sqrt{2}\cY^{\frac{3}{2}}
+256\cY^{\frac{7}{4}} +35\sqrt{2}\cY
\\
&-7\sqrt{2}\cY^{\frac{1}{2}}+5\sqrt{2}\big),
\\
&G_1(1)=\frac{256-144\sqrt{2}}{8192}\pi.
\end{aligned}
\end{equation}

\subsection{Phase shift}
\label{AppendixB}

We find the phase shift only in the case $\be=\dl$, i.e., for
$n=m=1$ in (\ref{eq6}). The reason is that in other cases the
calculations become too bulky and yield finally multiline
formulae. One more reason is that for other values of $m$ and $n$,
in the calculations we face with the problem of finding roots for
polynomials of high degrees and we can not find them explicitly.

We proceed to the calculations. First we compare formulae
(\ref{Denis20}) and (\ref{eq1}) and see that by (\ref{eq4})
\begin{equation*}
F(z)-B=-\frac{1}{\dl} g(z,+\infty)=\frac{1}{\dl} G(z).
\end{equation*}
Hence, it follows from (\ref{eq7}), (\ref{eq14a}) that
\begin{equation*}
F(z)-B=\frac{2^5}{\dl^3} \E^{4\dl z} G_1(\E^{2\dl
z})=\frac{\pi}{\dl^3} \frac{\E^{\dl z}(\E^{2\dl z}+4\E^{\dl
z}+1)}{(\E^{\dl z}+1)^4}.
\end{equation*}
We substitute this formula into (\ref{Denis19}) and make the
change of variable $y=\E^{\dl Z}$:
\begin{align}
&\Delta x=\pm\frac{1}{\dl} G_3\left(-\frac{M V_k \dl^4}{\pi
\epsilon}\right),\label{eq15}
\\
&G_3(\cZ):=\int\limits_{0}^{+\infty}
\frac{(y^2+4y+1)dy}{y(y^2+4y+1)-\cZ(y+1)^4}.\label{eq16}
\end{align}
It is easy to check the estimate
\begin{equation*}
0\leqslant\frac{y(y^2+4y+1)}{(y+1)^4}\leqslant\frac{3}{8},\quad
y\in[0,+\infty),
\end{equation*}

and thus, function $G_3(\cZ)$ is well-defined only for
$\cZ\in(-\infty,0)\cup(3/8,+\infty)$. In what follows we assume
that $\cZ=-\frac{M V_k\dl^4}{\pi \epsilon}$ ranges exactly in this
domain.

The integrand in (\ref{eq16}) is a rational function and to
integrate it, we need to expand it into the sum of simple
fractions. The latter requires the knowledge of the roots for the
denominator. Fortunately, we are able find these roots explicitly.
They are given by the identities:
\begin{equation}\label{eq22}
\begin{aligned}
y_1:=&-\frac{4\cZ-1+\sqrt{1+8\cZ}}{4\cZ}
\\
&+\frac{\sqrt{1+(4\cZ-1)\sqrt{1+8\cZ}}}{2\sqrt{2}\cZ},
\\
y_2:=&-\frac{4\cZ-1+\sqrt{1+8\cZ}}{4\cZ}
\\
&-\frac{\sqrt{1+(4\cZ-1)\sqrt{1+8\cZ}}}{2\sqrt{2}\cZ},
\\
y_3:=&-\frac{4\cZ-1-\sqrt{1+8\cZ}}{4\cZ}
\\
&+\frac{\sqrt{1-(4\cZ-1)\sqrt{1+8\cZ}}}{2\sqrt{2}\cZ},
\\
y_4:=&-\frac{4\cZ-1-\sqrt{1+8\cZ}}{4\cZ}
\\
&-\frac{\sqrt{1-(4\cZ-1)\sqrt{1+8\cZ}}}{2\sqrt{2}\cZ}.
\end{aligned}
\end{equation}

Generally speaking, there roots are complex-valued and this is why
we fix the branch of the square root by the requirement
$\sqrt{\E^{\iu\theta}}=\E^{\iu\frac{\theta}{2}}$,
$\theta\in(-\pi,\pi]$.

We consider first the case $1+8\cZ<0$, i.e.,
$\cZ\in(-\infty,-1/8)$. Here all the roots are complex-valued.
Roots $y_1$ and $y_3$ are complex-conjugate and the same is true
for $y_2$, $y_4$:
\begin{equation}\label{eq17}
\begin{aligned}
&y_1=P_1+\iu Q_1,\quad y_2=P_2+\iu Q_2,
\\
& y_3=P_1-\iu Q_1,\quad y_4=P_2-\iu Q_2.
\end{aligned}
\end{equation}
The denominator in (\ref{eq16}) obeys the representation:
\begin{align*}
y(y^2+4y+1)&-\cZ(y+1)^4
\\
&=-\cZ\big((y-P_1)^2+Q_1^2\big) \big((y-P_2)^2+Q_2^2\big).
\end{align*}
By means of this representation it is straightforward to make sure
that
\begin{align*}
\frac{(y^2+4y+1)dy}{y(y^2+4y+1)-\cZ(y+1)^4}=  &-\frac{1}{\cZ}\Bigg( \frac{K_1(y-P_1)+K_2}{(y-P_1)^2+Q_1^2} \\
&- \frac{K_1(y-P_2)+K_3}{(y-P_2)^2+Q_2^2}\Bigg),
\end{align*}
where
\begin{equation}
\begin{aligned}
K_1:=&\frac{K_3-K_2+1}{P_1-P_2},
\\
K_2:=&\Big(\big(Q_2^2+Q_1^2+(P_2-P_1)^2\big)(P_1^2+4P_1+1)
\\
&+Q_1^2\big(P_1^2+Q_1^2-P_2^2-Q_2^2-2P_1P_2-8P_2-2\big)\Big)
\\
&/\Big((P_2^2+Q_2^2-P_1^2-Q_1^2)^2
\\
&-4(P_1-P_2)(P_2(P_1^2+Q_1^2)-P_1(P_2^2+Q_2^2))\Big),
\\
K_3:=&-\Big(\big(Q_2^2+Q_1^2+(P_2-P_1)^2\big)(P_2^2+4P_2+1)
\\
&+Q_2^2\big(P_2^2+Q_2^2-P_1^2-Q_1^2-2P_1P_2-8P_1-2\big)\Big)
\\
&/\Big((P_2^2+Q_2^2-P_1^2-Q_1^2)^2
\\
&-4(P_1-P_2)(P_2(P_1^2+Q_1^2)-P_1(P_2^2+Q_2^2))\Big).
\end{aligned}\label{eq18}
\end{equation}
Now we employ obvious formulae
\begin{align*}
&\int\limits
\left(\frac{K_1(y-P_1)+K_2}{(y-P_1)^2+Q_1^2}-\frac{K_1(y-P_2)+K_3}{(y-P_2)^2+Q_2^2}\right)
dy
\\
&=\frac{K_1}{2} \ln \frac{(y-P_1)^2+Q_1^2}{(y-P_2)^2+Q_2^2}
 \\
&\hphantom{=}+ \frac{K_2}{Q_1} \arctan \frac{y-P_1}{Q_1} -
\frac{K_3}{Q_2} \arctan \frac{y-P_2}{Q_2}
\end{align*}
to obtain
\begin{align*}
G_3(\cZ)=&\frac{K_1}{2\cZ} \ln \frac{P_1^2+Q_1^2}{P_2^2+Q_2^2} -
\frac{\pi}{2\cZ} \left(\frac{K_2}{|Q_1|}-\frac{K_3}{|Q_2|}\right)
\\
&- \frac{K_2}{Q_1\cZ}\arctan\frac{P_1}{Q_1}
+\frac{K_3}{Q_2\cZ}\arctan\frac{P_2}{Q_2}.
\end{align*}
This formula, (\ref{eq18}), (\ref{eq22}), (\ref{eq17}),
(\ref{eq16}), and (\ref{eq15}) provide the final expression for
the phase shift once $\frac{M V_k \dl^4}{\pi
\epsilon}>\frac{1}{8}$.

We proceed to the case $\cZ\in(-1/8,0)$. Here all the roots
$y_1,\ldots,y_4$ are real and negative. The denominator in
(\ref{eq16}) can be represented as
\begin{equation*}
y(y^2+4y+1)-\cZ(y+1)^4=-\cZ(y-y_1)(y-y_2)(y-y_3)(y-y_4).
\end{equation*}
The integrand in (\ref{eq16}) can be expanded as
\begin{align*}
\frac{y^2+4y+1}{y(y^2+4y+1)-\cZ(y+1)^4}=-\frac{1}{\cZ}
\Bigg(&\frac{R_1}{y-y_1}+ \frac{R_2}{y-y_2}
\\
&+ \frac{R_3}{y-y_3}+ \frac{R_4}{y-y_4}\Bigg),
\end{align*}
where
\begin{equation}\label{eq19}
\begin{aligned}
&R_1:=\frac{y_1^2+4y_1+1}{(y_1-y_2)(y_1-y_3)(y_1-y4)},
\\
&R_2:=\frac{y_2^2+4y_2+1}{(y_2-y_1)(y_2-y_3)(y_2-y4)},
\\
&R_3:=\frac{y_3^2+4y_3+1}{(y_3-y_1)(y_3-y_2)(y_3-y4)},
\\
&R_4:=\frac{y_4^2+4y_4+1}{(y_4-y_1)(y_4-y_2)(y_4-y3)}.
\end{aligned}
\end{equation}
Integrating the obtained identity, we get
\begin{equation}\label{eq1919}
G_3(\cZ)=\frac{1}{\cZ}\sum\limits_{j=1}^{4}R_j\ln|y_j|.
\end{equation}
Together with (\ref{eq15}), (\ref{eq16}), (\ref{eq22}),
(\ref{eq19}) it leads us to the formula for the phase shift as
$0<\frac{M V_k \dl^4}{\pi \epsilon}<\frac{1}{8}$.

It remains to study the case $\cZ\in(3/8,+\infty)$. Here the roots
$y_1$, $y_2$ are real and negative, while $y_3$, $y_4$ are
complex-valued and complex conjugate:
\begin{equation}\label{eq20}
y_3=P_3+\iu Q_3,\quad y_4=P_3-\iu Q_3.
\end{equation}
The representation for the denominator in (\ref{eq16}) reads as
\begin{equation*}
y(y^2+4y+1)-\cZ(y+1)^4=-\cZ(y-y_1)(y-y_2)\big((y-P_3)^2+Q_3^2\big).
\end{equation*}
The integrand in (\ref{eq16}) satisfies the identity
\begin{align*}
\frac{y^2+4y+1}{y(y^2+4y+1)-\cZ(y+1)^4}=&-\frac{1}{\cZ}
\Bigg(\frac{L_1}{y-y_1}+ \frac{L_2}{y-y_2}
\\
&- \frac{(L_1+L_2)(y-P_3)+L_3}{(y-P_3)^2+Q_3^2}\Bigg),
\end{align*}
where
\begin{equation}\label{eq23}
\begin{aligned}
L_1:=&\frac{y_1^2+4y_1+1}{\big((P_3-y_1)^2+Q_3^2\big)(y_1-y_2)},
\\
L_2:=&\frac{y_2^2+4y_2+1}{\big((P_3-y_2)^2+Q_3^2\big)(y_2-y_1)},
\\
L_3:=&-P_3(L_1+L_2)+y_1 L_1 + y_2 L_2-1.
\end{aligned}
\end{equation}
Integrating the above formulae, we arrive at
\begin{align*}
G_3(\cZ)=\frac{1}{\cZ}\Bigg(&L_1\ln|y_1|+L_2\ln|y_2|
\\
&-\frac{L_1+L_2}{2}\ln (P_3^2+Q_3^2)
\\
&+ \frac{\pi L_3}{2|Q_3|}  + \frac{L_3}{Q_3}
\arctan\frac{P_3}{Q_3}\Bigg).
\end{align*}
This formula, (\ref{eq15}), (\ref{eq16}), (\ref{eq22}),
(\ref{eq20}), (\ref{eq23}) yield the analytic expression for the
phase shift once $\frac{M V_k \dl^4}{\pi \epsilon}<-\frac{3}{8}$.

\end{document}